\documentclass[aps,prb,preprint,a4paper,unsortedaddress,amsmath,amssymb,byrevtex,floatfix]{revtex4}

\usepackage{amsmath}
\usepackage{amssymb}
\usepackage[latin1]{inputenc}
\usepackage{graphicx}
\usepackage{times}

\begin{document}


\title{On the exposure to mobile phone radiation in trains}

\author{J. Ferrer}
\author{L. Fernández-Seivane}
\affiliation{Departamento de F\'{\i}sica, Universidad de Oviedo, 33007 Oviedo, Spain}
\author{J. M. Hernando}
\affiliation{Departamento de Se\~nales, Sistemas y Radiocomunicaciones, 
             Universidad Politécnica de Madrid, 28040 Madrid, Spain}
\author{M. B. Castán}
\author{L. García}
\author{J. M. Vázquez}
\affiliation{Departamento de Tecnologías de la Red, Telefónica Móviles España,
28224 Madrid, Spain}


\begin{abstract}
This report presents theoretical estimates of the Power Density levels which may be reached
inside trains. Two possible sources of high levels of radiation are discussed. The first one
arises since the walls of the wagons are metallic and therefore bounce back almost all
radiation impinging on them. The second is due to the simultaneous emission of a seemingly large 
number of nearby telephones. The theoretical study presented here shows that Power Densities stay 
at values below reference levels always.

\end{abstract}

\date{\today}

\maketitle 

Development in our society is tied up to the increasing usage of devices that generate
Electromagnetic Fields (EF). Mobile telephony has indeed become an integral part of modern 
style of life. 
But, while mobile handsets have made life easier and more pleasant, concern has also grown as to the
possible adverse consequences for human and animal health of short and long-term exposure to 
their emissions.

Absorption of radiation in the frequency range of mobile telephones is determined by the 
Specific Absorption Rate (SAR), which is measured in Watt/Kg. The International Commission 
on Non-Ionizing radiation (ICNIRP) has set a basic restriction on SAR for 
the body parts of the general public of 2 Watt/Kg. \cite{ICNIRP} For practical purposes, ICNIRP has  
established a Power Density $\mathbb{S}$ in air of $\nu/200 \,$ Watt/m$^2$ as an alternative 
reference level, where $\nu$ is the frequency of the EF measured in MHz. Since mobile telephone 
antennas emit radiation at either 0.9 or 1.8 GHz their
reference levels are of 4.5 or 9 Watt/m$^2$, respectively. Compliance with the reference level 
ensures compliance with the basic restriction. 
Additionally, since a substantial amount of the radiation emitted by a mobile handset is
absorbed by its own user, information of the SAR at the head is disclosed for all 
handsets made by members of the so-called Mobile Manufacturers Forum. \cite{MMF} Typical values, 
which range between 0.5 and 1.1 Watt/Kg, are always smaller than ICNIRP basic restriction.

There has been some recent controversy as to whether Power Densities far exceeding ICNIRP reference 
levels could be attained in the interior of train wagons, whose metallic case prevents leakage of the 
radiation emitted by the handsets to the outside world.  
Some authors have argued that the wagon windows would not suffice to
dissipate the radiation and, therefore, large levels of 
radiation would be absorbed both by the user and by other passive-user travelers, with potentially 
damaging health effects. \cite{Hon02,Hon03} If confirmed, such predictions would lead to a 
serious public health issue, since millions of persons pick either commuter or long-distance trains 
every day, and a substantial percentage of them use their mobile phones during their trips. 

We address in this article two possible coexisting scenarios of high levels of 
radiation. The {\it Proximity scenario} is due to the fact that it is
plausible that a passenger be in close proximity to a large number of telephones in use,
due to the small dimensions of a wagon, thereby receiving radiation from all those handsets.
The {\it Sauna scenario} corresponds to a situation where the level of ambient radiation 
inside the wagon cavity, $\mathbb{S}_{sauna}$, is large due to a very low capability of drains 
to suck radiation out. In contrast, we find that $\mathbb{S}_{sauna}$ is small, since it is 
proportional to the effective output power of all handsets, ${\cal P}$, 
divided by the total effective area of all dissipating surfaces, $S_{eff}$
\begin{equation}
\mathbb{S}_{sauna}=\frac{{\cal P}}{S_{eff}}\label{PD_eq}
\end{equation}
This equation may be interpreted as a sort of generalized Gauss' law, which takes account of the 
partial transparency of the boundaries. Not only windows, but also passengers take their share in 
$S_{eff}$, making it eventually a large denominator, and reducing the ambient levels of 
$\mathbb{S}_{sauna}$. 
This can be understood from a common experience with own's microwave oven: it takes 
always a considerably shorter amount of time to warm up one cup of coffee than, say, three or four.

Power Density distributions inside train wagons vary widely depending on both the shape of the wagon 
and windows, and on the number and position of passengers and emitting handsets. 
We could therefore perform a large number of simulations of the possible configurations of sources and
drains of EF, followed by
the adequate statistical analysis of the obtained data. But we believe that a better option is to set 
a qualitative description that focuses on the general laws governing the physical behavior
of the radiation field in the train.

We discuss the Sauna scenario first.
We model the wagon as a metallic cavity whose windows are covered by glass. We call $V$ the 
total volume enclosed by such a case, and $S_M$, $S_W$, the surface areas of the metallic walls and the 
windows. The floor of the wagon is covered by a rubber sheet of area $S_F$. The wagon is populated by 
$C$ seats and $D$ passengers, each of which has an area $s_C$, $s_D$. There are $H$ handsets inside the wagon,
all emitting radiation at the same frequency $\nu$. We call their total output power, averaged over
time ${\cal P}_H$, and ${\cal P}_0={\cal P}_H/H$ the output power per handset. There exists a residual 
radiation coming from the outside world, whose power is ${\cal P}_{res}$. 

We therefore have a cavity filled with microwave radiation in a single mode of frequency $\nu$. That is to say,
the cavity is filled with a large number $N$ of photons of frequency $\nu$. 
The basic rules of equilibrium Statistical Physics say that the energy of the field is 
$U=h\,\nu\,(N+1/2)$. Then, the Power Density can be expressed as
\begin{equation}
\mathbb{S}_{sauna}=\frac{c\,U}{V}\approx \frac{h\,\nu\,c}{V}\,N
\end{equation}  

The number of photons in the cavity in the stationary state $N_{st}$ will be a balance 
between those fed inside it and those lost at its surfaces,   
\begin{equation}
\frac{dN}{dt}=\frac{dN_{in}}{dt}-\frac{dN_{out}}{dt}=0\label{Balance}
\end{equation}

The rate of photons that are poured in by the handsets or through the windows is
\begin{equation}
{\cal P}={\cal P}_H+{\cal P}_{res}=h\,\nu\,\frac{dN_{in}}{dt}
\end{equation}

The number of photons that disappear through all the lossy surfaces may be estimated by the following
simple reasoning. We first notice that photons propagate in all directions. We therefore define the current 
density of photons at a given angle $ \hat{u}=(\theta,\phi) $ and at any point in the cavity as
\begin{equation}
\frac{d\vec{\it \j}}{d\Omega}=\frac{c\,N}{V}\,\frac{\hat{u}}{4\,\pi}
\end{equation} 

We assume that a beam of photons traveling along direction $\hat{u}$ impinges on a lossy surface $S$.
We call $t(\theta)$ the rate of photons that are lost either 
by absorption within the obstacle, or by transmission to the outside world. The total number of photons 
that are lost per unit solid angle and time, after impinging on the surface $S$ is
\begin{eqnarray}
\frac{d\dot{N}(\hat{u})}{d\Omega} &=&\int_S d\vec{S}\cdot\frac{d\vec{\it \j}}{d\Omega}\,\,t(\theta)
\end{eqnarray}
and the total number of photons lost per unit time at such surface may be estimated as
\begin{equation}
\dot{N}_{out}=\int_0^{\pi/2}\,d\theta\int_0^{2\,\pi}\,d\phi \,\,
\frac{d\dot{N}(\vec{u})}{d\Omega}\approx\,\frac{c\,S\,T}{V}\,N
\end{equation}
where $T$ is the angle- and surface-averaged coefficient $t(\theta)$.

The total number of photons lost through all surfaces is
\begin{equation}
\frac{dN_{out}}{dt}=\frac{c\,S_{eff}}{V}\,N
\end{equation}
where the effective surface
\begin{equation}
S_{eff}=S_M\,T_M+S_W\,T_W+S_F\,T_F+C\,s_C\,T_C+D\,s_D\,T_D
\end{equation}
takes account of the partial transparency of each specific surface through its averaged
lossy coefficient $T_i$. 

Eq. (\ref{Balance}) provides then with the number of photons in stationary situations
\begin{equation}
N_{st}= \frac{V}{h\,c\,\nu}\,\frac{{\cal P}}{S_{eff}}
\end{equation}
from which we can compute the Power Density, to find Eq. (\ref{PD_eq}).
We may also give an estimate of the average electric field inside the case, 
since the total energy of the EF is roughly proportional to its square,
\begin{equation}
U\approx h\,\nu\,N_{st}\approx\frac{V}{120\,\pi\,c}\,\,E^2
\end{equation}
whereby
\begin{equation}
E\approx \sqrt{\frac{120\,\pi\,{\cal P}}{S_{eff}}} \label{E_eq}
\end{equation}

Eq. (\ref{PD_eq}) provides with a qualitative estimate of the Power Density in terms of a few physical 
quantities which can be simply measured, such as surface areas, numbers of handsets and passengers and 
lossy coefficients. 
It is actually not that simple to determine accurately the lossy coefficients $T_i$, but we believe it 
worthy to make here an educated estimate of them. \cite{Wolf} For practical purposes, $T_M$ 
for aluminum may be set to zero, so that the metallic term drops out of the equation. The lossy 
coefficient of glass, which makes up the windows, is approximately equal to one, and therefore, 
$T_W\approx 1$. The conductivity of rubber at microwave frequencies is of order 1 sec$^{-1}$, in 
Gaussian units. Then we find that $T_F \approx 0$. Since the metallic parts of seats do not matter, we 
only need to find the absorption coefficient of wool, or similar manufactures, which is again very 
small; then $T_C\approx 0$, and the seats term also drops out. We finally assume that reflection of EF 
at each passenger mostly occurs at the skin, and use its conductivity 
$\sigma\approx5\times 10^9$ sec$^{-1}$ to find $T_D\approx 0.5$. \cite{Dim94} Such estimates allow us 
to provide a simpler version of Eq. (\ref{PD_eq}),
\begin{equation}
\mathbb{S}_{sauna}\approx \frac{{\cal P}}{S_W+D\,T_D\,s_D}\label{PD_simple}
\end{equation}

As a reference, the average surface area of a passenger ranges from 1.5 to 2 m$^2$, and $S_W$ is of 
about 30-40 m$^2$ for many train wagons. Eq. (\ref{PD_simple}) can then be used to make rough 
estimates of $\mathbb{S}_{sauna}$ absorbed by passengers and benchmark them against references levels 
supplied by 
ICNIRP or other institutions. We find that $\mathbb{S}_{sauna}$ saturates to 
${\cal P}/T_D\,D\,s_D$, when 
the surface area of passengers is much larger than that of windows. Such Power Density is much smaller 
than the 
reference levels of ICNIRP even when all passengers in the train are using simulateneously one handset. 

A previous estimate of this magnitude, performed by Hondou\cite{Hon02} yielded a Power Density 
$\mathbb{S}_{sauna}={\cal P}/ S_W$. Hondou assumed that the radiation emitted by all handsets had to be 
absorbed by each individual passenger, therefore predicting pretty alarming
levels of radiation.  Fig. \ref{Fig1} indeed shows how his predictions for $\mathbb{S}_{sauna}$ exceed our 
estimates by at least one order of magnitude.

The most crude simplification in the model is the assumption that radiation is emitted at a 
constant pace, in an homogeneous and isotropic fashion throughout the whole wagon, and in a single 
mode of frequency $\nu$. A more accurate handling of the sources of radiation and dissipation should 
only lead to a quantitative correction to the output power as long as the number of active handsets is 
large enough. 

We now turn to describe the proximity scenario. We suppose that there is a shell of several active 
handsets $G$ placed at a close distance $d$ around a probe. We moreover assume that their antennas are 
directly oriented towards the probe. Then, a straightforward application of Gauss's theorem yields a 
Power Density
\begin{equation}
\mathbb{S}_{prox}=\frac{A\,G\,{\cal P}_0}{4\,\pi\,d^2}
\end{equation}
where $A=1$ if the radiation pattern is isotropic or some number of order 5/4 if it is dipolar. 
We notice that the contribution from a second shell is much smaller than the previous estimate and can 
therefore be discarded. Kramer and coworkers, who  have also discussed the proximity effect, 
reached similar conclusions. \cite{Kra02}

The Power Density at the head of a mobile phone user might be estimated by assuming that 
the handset is usually placed side by side to an ear. A simple calculation then shows that about a 
third of the output power ${\cal P}_0$ is directed towards the head, a value roughly 
consistent with the numerical simulations by Dimbylow and coworkers. \cite{Dim94} The average Power
Density coming from this source is, accordingly,
\begin{equation}
\mathbb{S}_{head}=\frac{A\,{\cal P}_0}{3\,S_{head}}
\end{equation}
where $S_{head}$ is the area of the zones in the head hit by the radiation.

The total Power Density is, to conclude, the sum of the three discussed contributions
\begin{equation}
\mathbb{S}_T=\mathbb{S}_{sauna}+\mathbb{S}_{prox}+\mathbb{S}_{head}
\end{equation}
We now take a worst case scenario to place an upper bound on $\mathbb{S}_T$ in an actual train. We  
choose a small wagon with a floor area of 35 m$^2$, that is populated by 300 passengers. Such packed
situations have been argued to occur in some commuter trains in Japan. \cite{Hon02} The 
average distance among passengers is therefore of 35 cm. We also assume that all of them are using a 
handset at the same time. We take ${\cal P}_0 = 0.25$ Watt, that corresponds to the 
time-averaged output power of a GSM-900 handset, whose peak output power of about 2 Watt is distributed
among 8 channels. We finally assume that $S_{head}\sim 0.05$ m$^2$.

Then, the most important sources of radiation are the direct exposure to own's telephone
and that due to the Proximity scenario, each providing $\mathbb{S}_{prox,head} \sim 1.5$ Watt/m$^2$. The Sauna effect
provides a mere $\mathbb{S}_{sauna} \sim 0.3$ Watt/m$^2$, where we do not even take into account that only 
radiation poured out of the handsets towards the ceiling or the floor contributes to ${\cal P}$ in 
this case. The added contribution of the three sources lead in any case to a total Power Density smaller than ICNIRP 
reference levels, no matter the number of emitting handsets and the size of the windows. 

As a summary, we find that the Power Density levels in a train are always smaller than 
ICNIRP reference levels, no matter the number of passengers each wagon may contain and the number of
handsets in use.

\newpage
\bibliography{004523APL}

\newpage
\begin{figure}
\caption{\label{Fig1} Power Density $\mathbb{S}_{sauna}$ as a function of the number of passengers D, 
assuming that all
of them are using a telephone at the same time, and have an area $s_D=1.5$ m$^2$. We also suppose 
that $S_W=$ 30 m$^2$ and ${\cal P}_0=0.25$ Watt / m$^2$. 
Solid and dashed lines are a plot of Eq. (\ref{PD_eq}) and the $\mathbb{S}$ predicted by Hondou, respectively. }
\end{figure}

\clearpage

\begin{figure}
\unitlength1cm
\includegraphics[width=8.5cm,height=8.5cm]{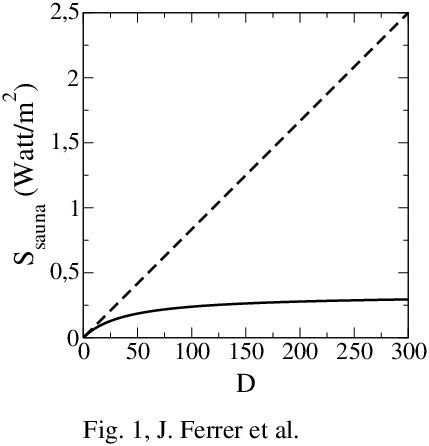}
\end{figure}

\end{document}